\documentstyle[amssymb,preprint,aps]{revtex}
\begin{document}
\title{Quasiparticle Relaxation Dynamics in Two Distinct Gap Structures: La$%
_{0.67}$Ca$_{0.33}$MnO$_{3}$ and LaMnO$_{3}$ }
\author{Y. H. Ren$^{1}$, H. B. Zhao$^{1}$, G. L\"{u}pke$^{1}$, Y. F. Hu$^{2}$%
, Qi Li$^{2}$, C. S. Hong$^{3\dagger }$ and N. H. Hur$^{3}$}
\address{$^{1}$Applied Science, the College of William \& Mary,\\
Williamsburg, VA 23187\\
$^{2}$Department of Physics, Pennsylvania State University, University Park,%
\\
PA 16802\\
$^{3}$Center for CMR Materials, Korea Research Institute of Standards and\\
Science, P. O. Box 102 Yusong, Daejeon, 305-600, Republic of Korea}
\date{\today}
\maketitle

\begin{abstract}
The spin and quasiparticle relaxation dynamics in La$_{0.67}$Ca$_{0.33}$MnO$%
_{3}$ and LaMnO$_{3}$ single crystals and thin films is investigated as a
function of temperature and magnetic field by pump-probe spectroscopy. The
unusually slow spin-lattice relaxation dynamics is governed by the
temperature- and magnetic field- dependent pseudogap in La$_{0.67}$Ca$%
_{0.33} $MnO$_{3}$, and by the temperature-independent Jahn-Teller gap in
LaMnO$_{3}$. Our results show that the coupled dynamics of charge, spin and
lattice is strongly correlated with the distinct gap structures in these
manganites.
\end{abstract}

\pacs{71.30.+h, 75.30.Fv, 78.47.+p }

Perovskite manganites, R$_{1-x}$A$_{x}$MnO$_{3}$ (R and A are rare- and
alkaline-earth ions, respectively) have attracted much attention because of
the large negative magnetoresistance and the magnetic-field-induced
phenomena. The strong correlation between spin, charge, orbital, and lattice
degrees of freedom leads to complex phase diagrams and to the coexistence of
various forms of ordering. Of particular interest in this context has been
the observation of a temperature-dependent pseudogap in a variety of
colossal magnetoresistance (CMR) oxides and its relationship to the
ferromagnetic (FM) metal-to-insulator (MI) transition \cite{Chuang
Science,Saitoh PRB,Park PRL,Dessau book}. Key to the pseudogap and many
other properties of the manganites appear to be the existence of nanoscale
charge/orbital fluctuations, which cooperate with Jahn-Teller distortions
and compete with the electron itineracy favored by double exchange \cite%
{Chuang Science}. A profound knowledge of these ordering processes and their
mutual interactions is essential for a better understanding of CMR.
Dynamical information on the electronic states of the pseudogap can be
obtained, if charge or spin excitations are introduced via pulsed
photoexcitation, and the relaxation dynamics of the quasiparticles (QP's) is
studied on a picosecond time scale. Ultrafast optical pump-probe
spectroscopy has provided significant insight in the dynamics of spin,
electron and lattice in metals \cite{Beaurepaire PRL,Koopmans PRL}, and
recently in transition metal oxides \cite{Dodge PRL,Kise PRL,Averitt PRL,Ren
JAP,Ren PRB,LiuPRB}.

Here, we report the time-resolved pump-probe reflectivity measurements of
photoexcited spin and QP relaxation dynamics in La$_{0.67}$Ca$_{0.33}$MnO$%
_{3}$ (LCMO) and LaMnO$_{3}$ (LMO) single crystals and thin films as a
function of temperature and magnetic field. Our results are fourfold: First,
both manganites reveal an unusually slow ($\sim $ 10 $\mu $s) carrier
relaxation process which disappears as the\ transition temperature is
approached from below. This slow decay process is attributed to spin-lattice
relaxation of carriers in localized (intragap) states. Second, the
quasiparticle relaxation rate is governed by the temperature-dependent
pseudogap in LCMO and by the temperature-independent Jahn-Teller distortion
gap in LMO. Third, the magnetic-field-induced increase of spectral weight in
the pseudogap is evident in both the quasiparticle and spin relaxation
dynamics in LCMO. Fourth, the polaron relaxation time, $\tau _{B}$, is
temperature-independent in LMO, whereas $\tau _{B}$ follows a $%
(T_{C}-T)^{-\beta }$\ dependence in LCMO, indicating that the quasiparticle
relaxation dynamics involves fully spin-aligned pseudogap states near the
Fermi energy $E_{F}$. Our results show for the first time that the coupled
dynamics of charge, spin and lattice is strongly correlated with the
distinct gap structures in these manganites.

LCMO and LMO single crystals and thin films were grown by the floating zone
method and pulsed laser deposition, the details of which are reported
elsewhere \cite{Ren PRB,QiLi APL,Hong CM}. The samples were characterized by
electrical resistivity and magnetization measurements. The LCMO single
crystal and 400-nm thin film have a Curie temperature $T_{C}$ = 225 K and
260 K, respectively. The LMO single crystal shows a Neel temperature $T_{N}$%
\ = 145 K. For the transient reflectivity measurements the samples were
mounted in an optical cryostat. The laser system consists of a Ti:sapphire
regenerative amplifier (Spitfire, Spectra-Physics) and an optical parametric
amplifier (OPA-800C, Spectra-Physics)\ delivering 100-fs short pulses at a
1-kHz repetition rate tunable from 600 nm to 10 $\mu $m. A two-color
pump-probe setup is employed with the pump beam power $<$ 6 mW and the probe
beam power $<$ 1 mW. The unfocused pump beam, spot-diameter $\sim $3 mm$^{2}$%
, and the time-delayed probe beam are overlapped on the sample with their
polarization perpendicular to each other. The reflected probe beam is
detected with a photodiode detector. A SR250 gated integrator \& boxcar
averager, and a lock-in amplifier are used to measure the transient
reflectivity change $\Delta R$ of the probe beam. For the magnetic-field
dependent measurement we used a 9-Tesla superconducting magnetic cryostat
from Oxford Instruments, Inc.

Figures 1 (a) and 1 (b) show the time evolution of $\Delta R$ from LCMO and
LMO single crystals at different temperatures. The pump and probe wavelength
is 800 nm. At low temperature (T $<$$<$ T$_{C}$, T$_{N}$), $\Delta R$ shows
initially a fast biexponential decay with relaxation times $\tau _{A}$= 0.5 $%
\backsim $ 4 ps and $\tau _{B}$ = 50 $\sim $ 100 ps. The fast process
reveals the thermalization of photoexcited quasiparticles which occurs on a
time scale $\tau _{QP}$ = 0.3 $\backsim $ 3 ps \cite{Han PRL}. The second
process is characteristic for polaron relaxation \cite{Ren PRB,Y Zhao PRL}.

In addition to the initial fast relaxation processes, a very long-lived
negative $\Delta R$ signal remains sufficiently long, decay time $\tau
_{SL}\sim $ 10 $\mu $s, that a negative $\Delta R$ is clearly observable
even after 1 ms (Figs. 1(a) and 1(b)). This component is present in both the
metallic and insulating phase as observed in the transient optical
reflectivity and transmission measurements from charge-ordered Pr$_{0.67}$Ca$%
_{0.33}$MnO$_{3}$ (PCMO), La$_{0.67}$Sr$_{0.33}$MnO$_{3}$ (LSMO), and Nd$%
_{0.67}$Sr$_{0.33}$MnO$_{3}$ (NSMO). However, there is no evidence of this
component in the transient reflectivity change of the paramagnetic phase of
LSMO, which shows a phase transition from ferromagnetic metal to
paramagnetic metal at 325 K. Therefore, the long-lived negative component of 
$\Delta R$ is ascribed to a slow spin relaxation process of the magnetically
ordered phase.

Figure 2 illustrates the temperature dependence of $\Delta R$ measured at a
time-delay $\Delta t=500$ ps, referred to as $\Delta R^{\prime }$ in LCMO\
and LMO. The temperature dependence of $\Delta R^{\prime }$, for LCMO single
crystal ($\lambda _{pump}=800$ $nm/\lambda _{probe}=800$ $nm$) and thin film
($\lambda _{pump}=800$ $nm/\lambda _{probe}=$ 5.2 $\mu m$) are qualitatively
the same and independent of wavelength (Fig. 2a). $\Delta R^{\prime }$\
increases with increasing temperature followed by an abrupt drop to zero
around $T_{C}$. In contrast, for LMO ($\lambda _{pump}=800$ $nm/\lambda
_{probe}=800$ $nm$)\ $\Delta R^{\prime }$ decreases slowly at low
temperature and drops toward zero abruptly around $T_{N}$ (Fig. 2(b)).

The fact, that in both manganites $\Delta R^{\prime }$ vanishes above the
transition temperature, is a strong indication that $\Delta R^{\prime }$
arises: a) from quasiparticle excitation involving localized (intragap)
states, and b) that the decay of these metastable states involves a
spin-flip process. The lifetime $\tau _{SL}\sim $ 10 $\mu $s of these
metastable\ states is comparable with the spin-lattice relaxation time
measured by the $\mu $SR technique \cite{Heffner PRB}.

The solid line in Fig. 2(b) shows that $\Delta R^{\prime }$ in LMO follows
approximately a $(T_{N}-T)^{\alpha }$ dependence ($\alpha $ $\cong $ 0.5),
which is similar to the temperature dependence of the magnetization \cite%
{Heffner PRB}. As the temperature approaches $T_{N}$ and the magnetic order
decreases in LMO the lifetime of the metastable state decreases due to
enhanced spin scattering and $\Delta R^{\prime }$ starts to decrease. The
non-zero $\Delta R^{\prime }$\ above $T_{N}$ indicates that a residual
anti-ferromagnetic (AFM) order exists in LMO above the Neel temperature \cite%
{Heffner PRB}. In contrast to LMO, the $T$-dependence of $\Delta R^{\prime }$
is more complex in LCMO. The transient reflectivity first increases with
increasing temperature and then for $T>0.9T_{C}$\ follows a $%
(T_{C}-T)^{\beta }$ dependence with $\beta \approx $ 0.7 - 0.9 (solid line
in Fig. 2(a)). A similar power-law dependence of $\Delta R^{\prime }$ has
been observed in other CMR manganites, e.g. Nd$_{0.67}$Sr$_{0.33}$MnO$_{3}$ %
\cite{Ren JAP}.

A simple physical model accounts qualitatively for these observations. The
various processes giving rise to the photoinduced reflectivity change in
LCMO are depicted in Fig. 3 (inset). An ultrashort laser pulse first excites
electrons via interband transitions. In our experiments, the films absorbed
10$^{18}$ - 10$^{20}$ photons/cm$^{3}$ per pulse; comparable to the
charge-carrier density ($\sim $10$^{20}$-10$^{21\text{ }}$holes/cm$^{3}$) in
LCMO; hence, one expects significant electron excitation during ultrashort
pump pulse illumination. These hot electrons very rapidly release their
energy via electron-electron and electron-phonon collisions reaching QP
states near the Fermi energy (step 1). This generates spin waves (magnons) %
\cite{Dodge PRL,LiuPRB}. Spin waves lead to the excitation of metastable
states with $\mu $s time scale \cite{Dodge PRL,LiuPRB}. The QPs can
recombine by interaction with states in the pseudogap (step 2) or relax to
metastable states via spin-flip processes (magnons are released) caused by
strong electron-lattice coupling (step 3). The carriers in the metastable
states will relax with a recombination rate $\gamma =1/\tau _{SL}$, while
magnons are absorbed (step 4). As the pseudogap opens up with increasing
temperature and the spectral weight of states at the Fermi level decreases
the decay rate of excited QPs decreases (step 2) and more quasiparticles
will be scattered into metastable states (step 3). This effect leads to the
initial rise of $\Delta R^{\prime }$ which is inverse proportional to the
density of states in the pseudogap, i.e., for $T\leq 0.9T_{C}$, $\Delta
R^{\prime }\propto (T_{C}-T)^{-b}$ with $b$ $\approx $ 0.3 - 0.5 (solid line
in Fig. 2(a)). With further increasing temperature ($T>0.9T_{C}$),\ $\Delta
R^{\prime }$ drops significantly toward zero due to the increase of the
density of states and occupation of the down-spin conduction electrons,
which introduces magnetic disorder.

Figure 4 shows magnetic-field dependent measurements of $\Delta R$\ at 305 K
(well above $T_{C}$) as a function of time for LCMO thin film grown on NdGaO$%
_{3}$ (110) substrate. Magnetic fields are applied in the Faraday geometry,
generating an isothermal magnetic entropy change. The results of Fig. 4 are
consistent with our model. The quasiparticle relaxation dynamics shows
strong magnetic field dependence for $B$ ranging from 0.5 to 3 T. The
magnetic-field dependence of $\Delta R$ for LCMO is very similar to the
temperature dependence of the zero-field $\Delta R$ shown in Fig. 1(a). The
increase in the amplitude of $\Delta R^{\prime }$ is due to the increase of $%
T_{C}$ with applied magnetic filed; this results in an increase of magnetic
correlations (spatially and temporally) with enhanced spin alignments. The
magnetic correlations are strong enough to overcome thermal fluctuations at
the new transition temperature.

The model also correctly explains the temperature dependence of the polaron
relaxation time $\tau _{B}$ for LMO and LCMO single crystals (Fig. 3). In
LMO single crystal, $\tau _{B}$ is found to be completely temperature
independent below $T_{N}$ consistent with a static Jahn-Teller gap. For LCMO
single crystal, $\tau _{B}$ remains nearly constant below 0.45 $T_{C}$ and
starts to increase above 0.7 $T_{C}$. A similar $T$-dependence has been
reported in LCMO thin films \cite{Ren PRB}. The polaron relaxation time $%
\tau _{B}$ follows a $(T_{C}-T)^{-\beta }$ dependence, i.e., $\tau
_{B}\propto 1/\Delta R^{\prime }$, shown as a solid line in Fig. 3. This
result strongly supports our model that the quasiparticle relaxation
dynamics involves fully spin-aligned pseudogap states near $E_{F}$.

In summary, we investigated the spin and quasiparticle relaxation dynamics
in LCMO and LMO single crystals and thin films as a function of temperature
and magnetic field by time-resolved pump-probe spectroscopy. The
quasiparticle relaxation rate is governed by the temperature- and magnetic
field- dependent pseudogap in LCMO and by the temperature-independent
Jahn-Teller distortion gap in LMO. Both manganites exhibit metastable
(localized) states with lifetime $\tau _{SL}\sim $ 10 $\mu $s, which decay
via spin-lattice coupling. Our results show that the coupled dynamics of
charge, spin and lattice is strongly correlated with the distinct gap
structures in these manganites.

This work is supported in part by NSF through grants: DMR-0137322,
IMR-0114124 (WM), and DMR-9876266 (PSU), and the Petroleum Research Fund
(PSU).

$^{\dagger }$ Current address for C. S. Hong, Department of Chemistry, Korea
university, Anam-dong, Sungbuk-ku, Seoul 136-701, Republic of Korea

\bigskip

\begin{center}
Figure Captions:
\end{center}

Fig. 1. Transient reflectivity $\Delta R$\ of: a). LCMO and b). LMO single
crystals around $T_{C}$ (or $T_{N}$) at 800 nm. The dotted lines indicate
the zero line.

Fig. 2. Temperature dependence of $\Delta R$\ (normalized) at $\Delta t=500$
ps: a) LCMO single crystal ($\lambda _{pump}=800$ $nm/\lambda _{probe}=800$ $%
nm$) and 400-nm thin film ($\lambda _{pump}=800$ $nm/\lambda _{probe}=5.2$ $%
\mu m$), and b) LMO single crystal ($\lambda _{pump}=800$ $nm/\lambda
_{probe}=800$ $nm$). The solid lines indicate the power-law dependence.

Fig. 3. The relaxation time $\tau _{B}$ as a function of temperature $T$ for
LCMO and LMO single crystals. The solid line indicates the power-law
dependence. The inset depicts a schematic diagram of carrier
excitation/relaxation processes in LCMO. For details see text.

Fig. 4. Transient reflectivity $\Delta R$ from LCMO as a function of time
for applied fields ranging from 0.5 to 3 T at 305 K ($\lambda _{pump}$, $%
\lambda _{probe}=800$ $nm$). The dotted lines indicate the zero line.

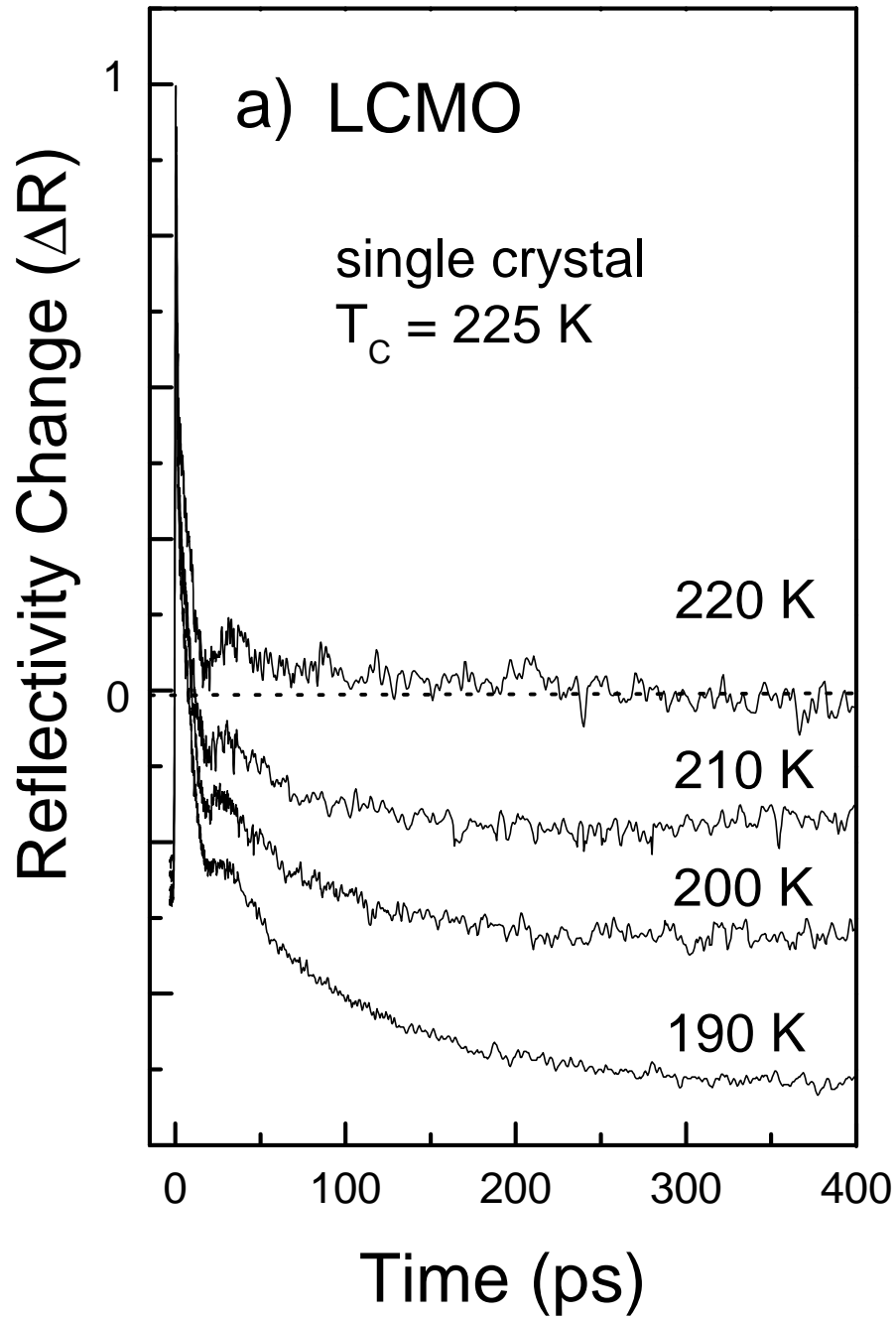
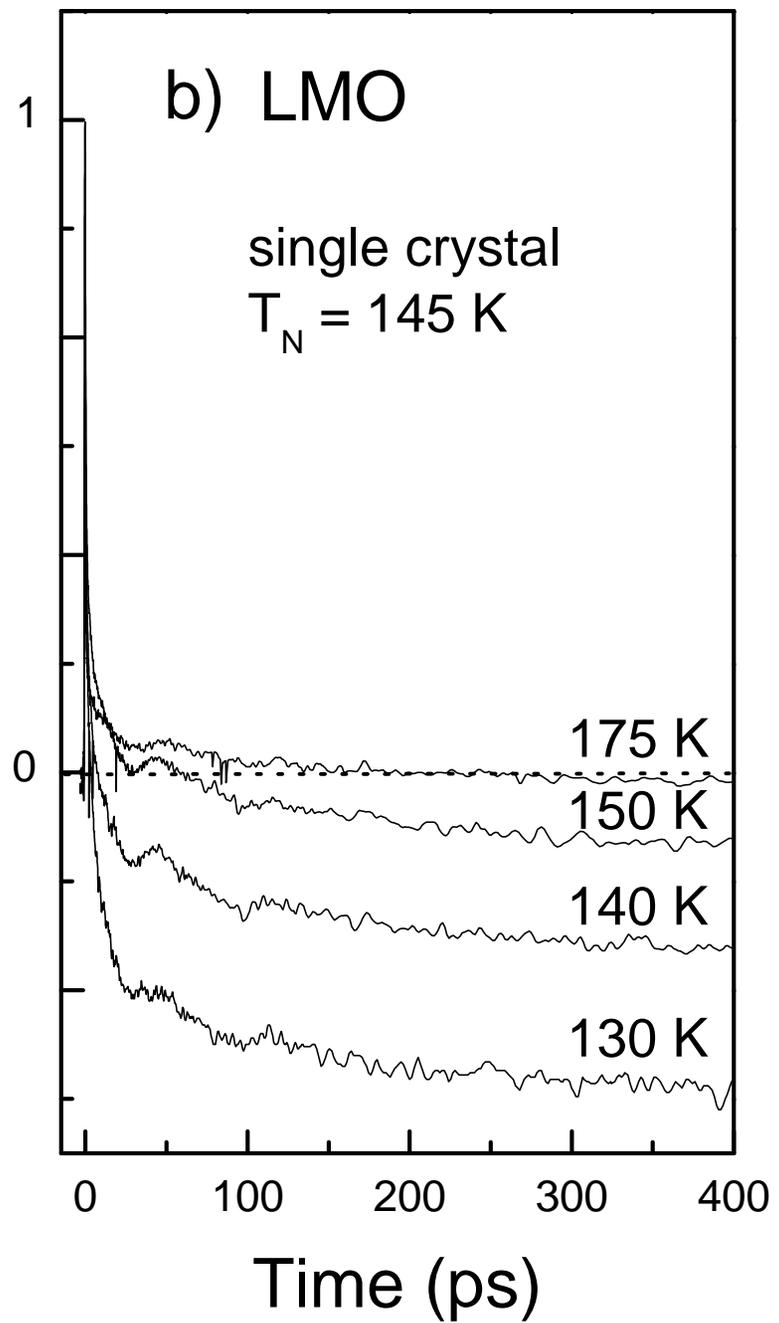
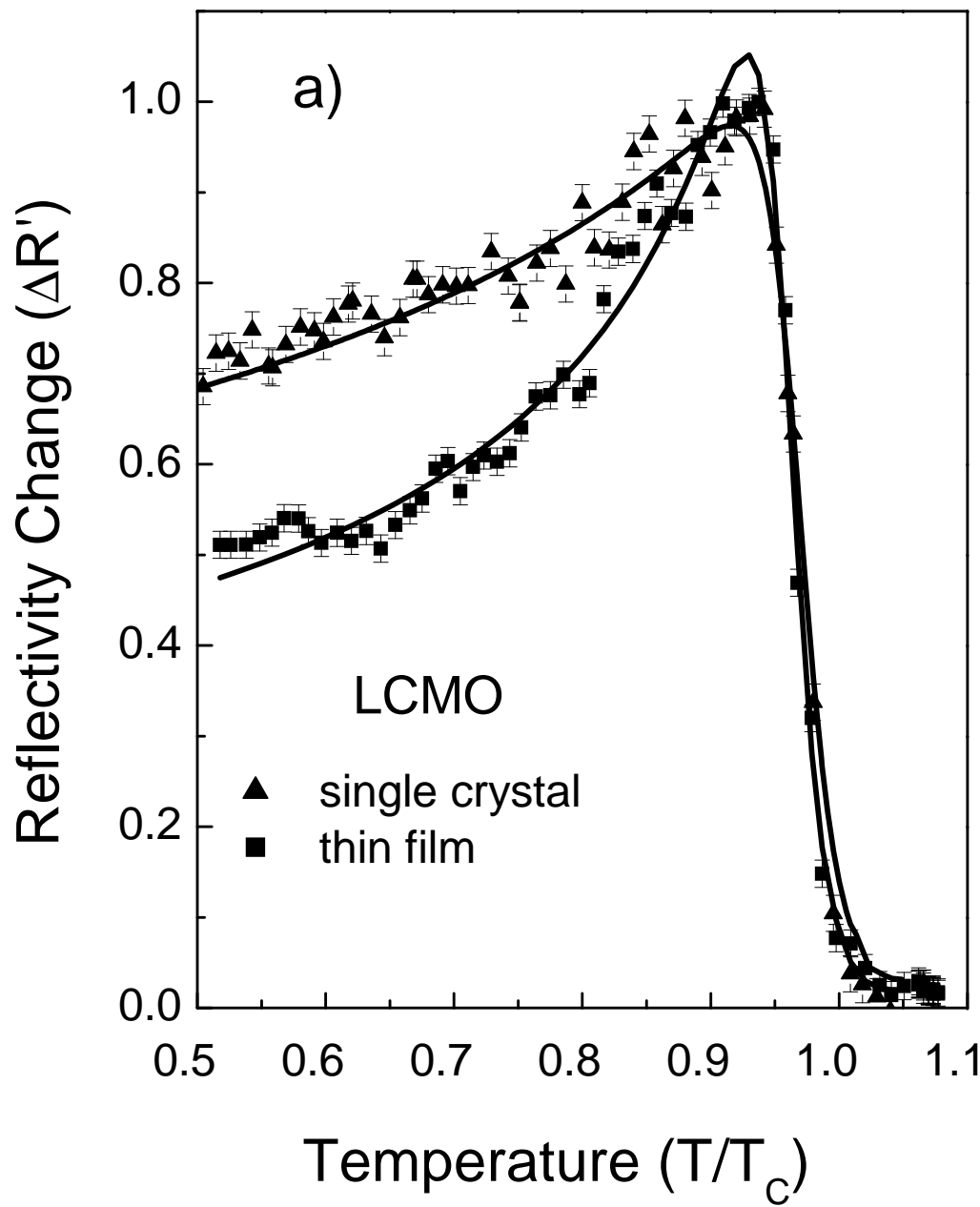
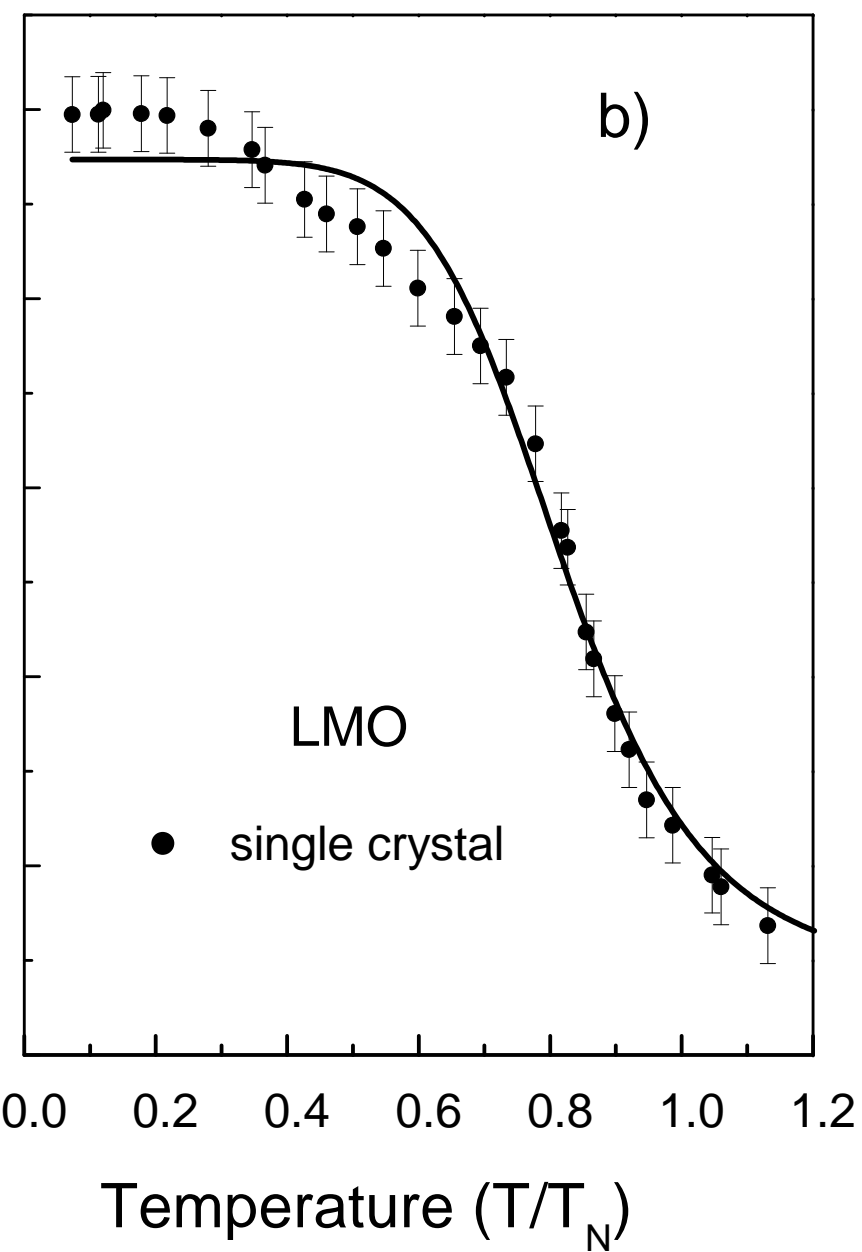
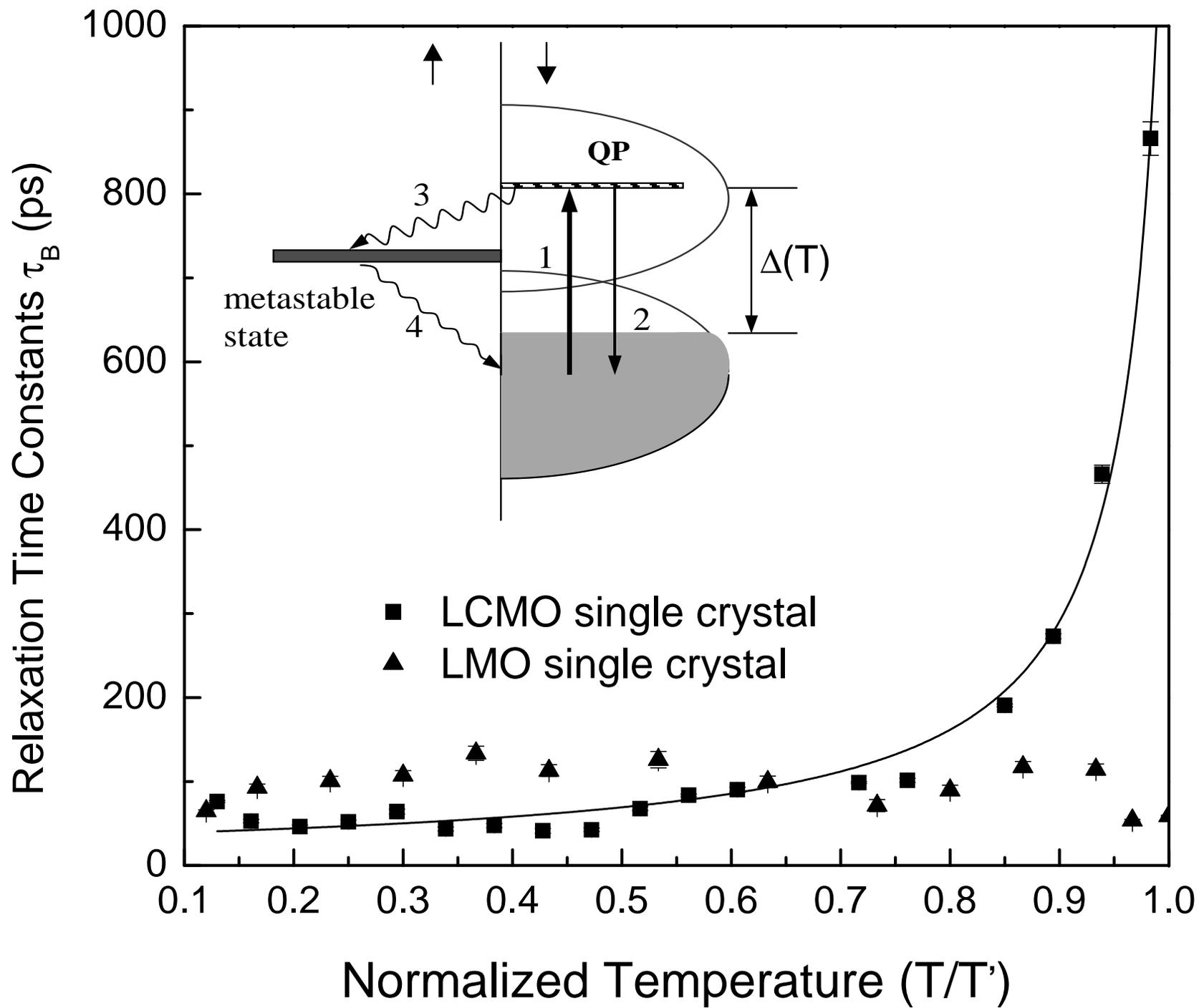
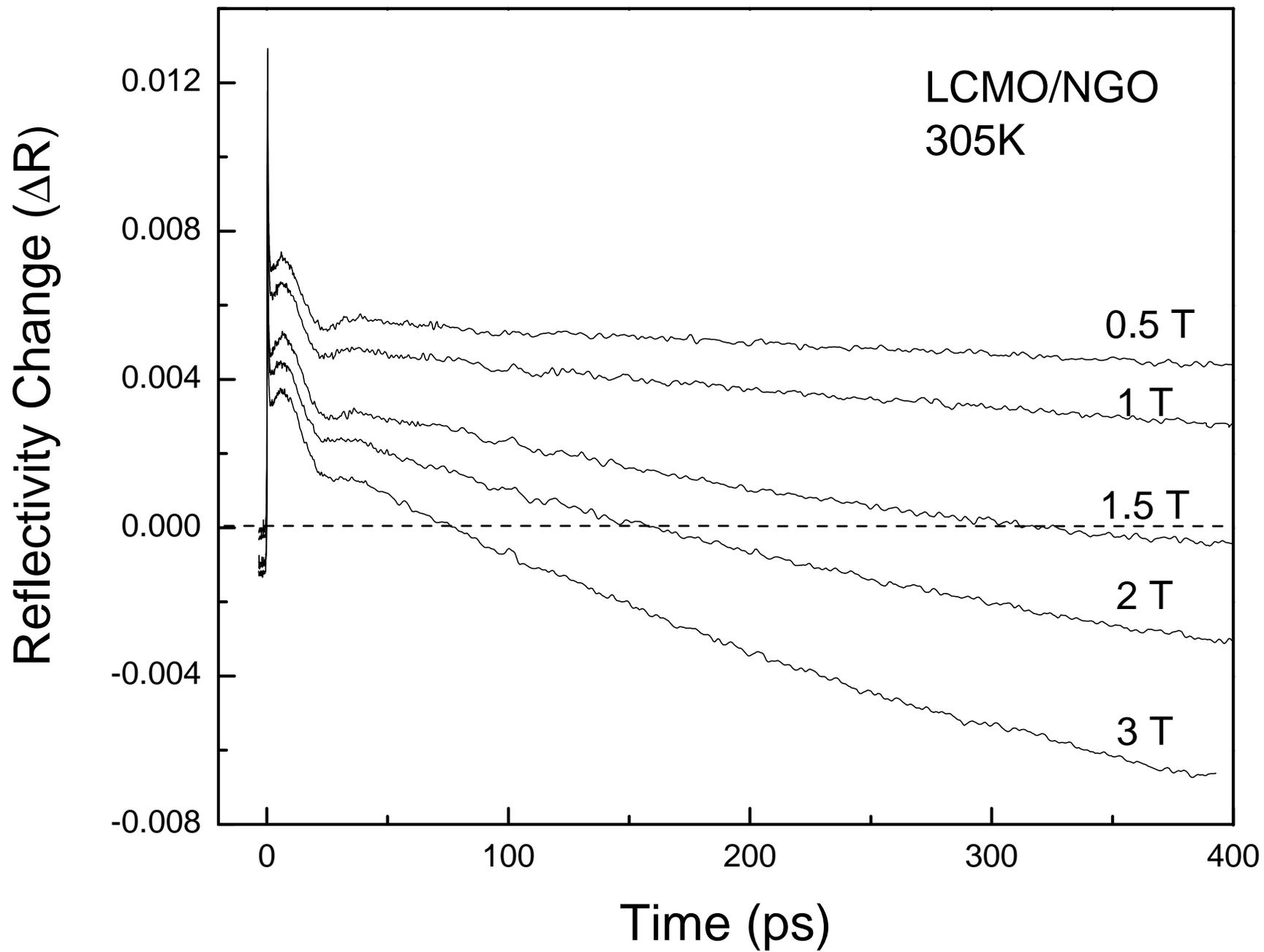
\end{document}